# The Soft X-ray Counterpart of Hanny's Voorwerp near IC 2497

*G. Fabbiano & M. Elvis*
*Center for Astrophysics | Harvard & Smithsonian, 60 Garden Street, Cambridge MA 02138, USA*

ABSTRACT

We report the detection in *Chandra* ACIS archival data of an elongated soft (< 3keV) X-ray feature to the south of the Compton Thick Active Galactic Nucleus (CT AGN) galaxy IC 2497, coincident with the emission line feature known as Hanny's Voorwerp. The data are consistent with the spatial correspondence between X-ray, optical emission line, and radio features detected in nearby obscured AGNs (e.g., ESO 428-G014). The X-ray luminosity of the (0.3 – 3.0 keV) soft feature is ~1.2 x $10^{40}$ erg s$^{-1}$. We infer an [OIII]/Soft-X-ray ratio in the range of ~ 200, consistent with the highest values measured in some of the clouds of NGC 4151. Overall, given the uncertainties, Hanny's Voorwerp appears to be a feature consistent with the ionization cone emission of nearby AGNs. We estimate an X-ray recombination time of ~2 x $10^7$ yr, longer than the [OIII] recombination time (~8000 yr). This suggests that extended soft X-ray components may be a better diagnostic of overall long-term activity, while detection of an [OIII] HV would point to a time-limited activity burst.

1. Introduction

The serendipitous discovery of an emission line feature (Hanny's Voorwerp), ~20 kpc south of the massive spiral galaxy IC 2497 (D~225 Mpc; 1" ~1 kpc; NED) led to the hypothesis of a past quasar outburst ~$10^5$ yr ago in this apparently quiescent galaxy (Lintott et al. 2009; Schawinski et al. 2010). This discovery led to follow up work, which established associated radio emission (Jozsa et al. 2009) and defined the detailed characteristics of the emission line feature (Keel et al. 2012). Using *Chandra*, soft, slightly extended (~1 kpc), X-ray emission was reported in IC 2497, with an emission temperature of ~1 keV, leading to the hypothesis of a hot bubble heated either by a low-activity nucleus over a long time, or as a result of past intense quasar activity (Sartori et al. 2016). A *Chandra-NuSTAR* combined spectral study (Sartori et al. 2018) showed that the nucleus of IC 2497 has the X-ray spectrum typical of a highly-obscured CT AGN with nuclear $N_H \sim 2 \times 10^{24}$ cm$^{-2}$, and current intrinsic luminosity $L_{bol} \sim 2$–$5 \times 10^{44}$ erg s$^{-1}$, a factor of ~50 lower than implied by the optical emission line properties of the Hanny's Voorwerp (HV).

As IC 2497 is a CT AGN, then it may have the more extended X-ray emission seen in many other obscured AGNs along the outflow axis (e.g., Bianchi et al. 2006, Levenson et al. 2006; Wang et al 2011a, b; Paggi et al 2012; Maksym et al. 2017, 2019; Fabbiano et al. 2018a). Moreover, the optical spectrum shows a strong Ne V line (Lintott et al. 2009), which requires soft X-rays (> 0.2 keV) for ionization. We therefore undertook a search for low surface brightness extended emission in the *Chandra* data, to compare its properties with those of other obscured nearby AGNs that our group has been studying in detail. In this paper we report our results.

## 2. Data Analysis and Results

For this study we have used the two archival *Chandra* ACIS-S Cycle 13 observations of IC 2497 (obsid 13966 and 14381; PI: Schawinski). This is the same data used in the previous *Chandra* work (Sartori et al. 2016, 2018). For this analysis, we used the latest version of CIAO (4.11), and the version of the display and analysis tool DS9 included in CIAO 4.11[1]. Following the CIAO threads, we merged the datasets, to obtain a merged event file with a total 112 ks exposure. Comparison with the parent images showed that no additional astrometric correction was needed. Given the relatively few detected counts, we used the ACIS-S native instrumental binning of the data for our analysis (0".492 pixel size. Note that the ACIS-S pixel size is larger than the Chandra PSF FWHM, ~0.2''.) For display purposes, we performed a simple Gaussian smoothing of the image, using the DS9 Gaussian smoothing function, with a radius of 5 image pixels and sigma of 2.5 pixels. The result is shown in Figure 1. Using wider smoothing functions (up to 4 pixel Gaussian) does not change the results.

An elongated feature is evident at low energies, extending ~20'' to the south of the IC 2497 central source, in the direction of HV. This feature is visible in the energy range (0.3-3.0 keV), corresponding to the soft spectral component of IC 2497 (Sartori et al 2018), but it is the clearest in the 0.3-1.5 keV band (Figure 1). Imaging in the 3-7 keV band results in a single point-like source at the position of the nucleus of IC 2497. This is the source responsible for the hard continuum and Fe K$\alpha$ emission of the CT AGN (Sartori et al 2018).

A comparison of the counts in the ten azimuthal sectors shown in Figure 1 (radii=12"-32") yields 44.3 net counts in the southern sector (0.3-1.5 keV), with a 5$\sigma$ statistical significance. The background counts were estimated from the other sectors, with the exclusion of the two sectors adjacent to that containing the feature. The data quality and analysis are similar to that in the original discovery observation of extended emission in NGC 4151 (Elvis et al. 1983). Assuming a thermal optically thin emission with kT~1 keV (consistent with the results of Sartori et al. 2016), and a line of sight[2] $N_H$=1.3 x $10^{20}$ cm$^{-2}$, the corresponding (0.3-3.0 keV) flux[3] is ~2 x $10^{-15}$ erg cm$^{-2}$ s$^{-1}$, corresponding to a luminosity of this feature of ~1.2 x $10^{40}$ erg s$^{-1}$.

---

[1] http://cxc.cfa.harvard.edu/ciao/

[2] Calculated with COLDEN; URL: http://cxc.harvard.edu/toolkit/colden.jsp
[3] Calculated with PIMMS; URL: http://cxc.harvard.edu/toolkit/pimms.jsp

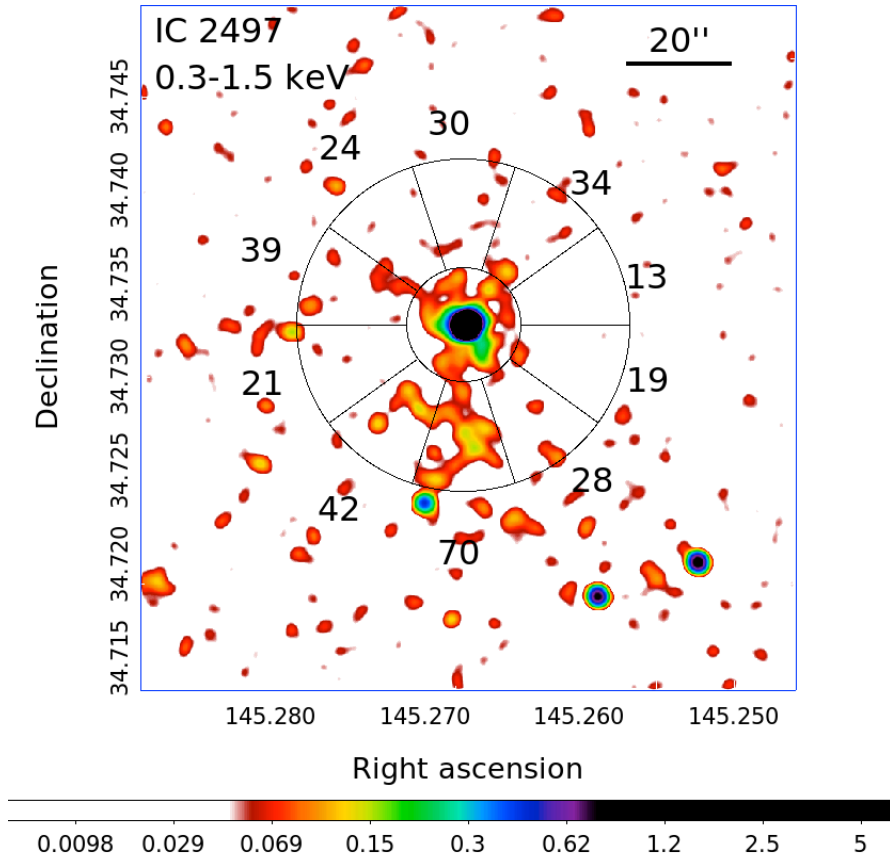

Figure 1. *Chandra* image of IC2497 in the 0.3 – 1.5 keV band. The data were gaussian smoothed over 5 ACIS pixels with a sigma of 2.5 pixels. The ring of azimuthal sectors (12''-32'') is centered on the CT nucleus. Numbers show the total number of counts in each sector of the image. The color scale at the bottom is in counts per pixel.

3. Discussion

Figure 2 shows the contours of the extended southern soft X-ray feature, overlaid onto the [OIII] *Hubble* image of HV from Sartori et al. (2016). The X-ray feature overlaps HV and appears, at low significance, to join it with the AGN nuclear source. This correspondence of soft extended X-ray emission and emission line features is typical of what we have observed in the ionization cones of nearby AGNs, as is the luminosity of the extended feature ~$10^{40}$ erg s$^{-1}$ (e.g., NGC 4151, Wang et al. 2011b; Mkn 573, Paggi et al 2012; ESO 428-G014, Fabbiano et al 2018a). The most similar case may be that of another probable "quasar relic", NGC 5252 (Dadina et al., 2010), which show similar soft X-ray and [OIII] spatial coincidence on a factor two smaller physical scale.

In these nearby AGNs we have also found correspondences between extended X-ray features (both soft and, in ESO 428-G014, also at energies > 3 keV) and radio jet emission. Interestingly, the same X-ray radio correspondence may occur in IC 2497, as shown by a comparison of the *Chandra* X-ray contours with Figure 2 of Jozsa et al. (2009). The 1.4 GHz radio continuum is also elongated in the N-S direction, overlapping the HV region, roughly in the same direction as the soft X-rays, reinforcing the similarity with nearby AGNs. Both radio continuum, optical emission line regions, and the *Chandra* emission to the south, reside in a local minimum of HI column density (see Figure 2 of Jozsa et al. 2009) in the clouds that are found both to the E and W of the radio continuum emission, at the redshift of IC 2497.

### 3.1 Inferred Physical Parameters

If the X-ray emission is mostly due to photoionization of a gas with the density inferred from the HI measurements (Jozsa et al. 2009), assuming the peak column density ($n_e$ ~$3\times10^{-3}$ = $N_H/D$ ~ $2\times10^{20}$cm$^{-2}$/~20 kpc), then its recombination time would be 2 x $10^7$ yr (see Wang et al 2010, Reynolds et al. 1995), much longer than the optical line recombination time of a few 1000 yr for the hydrogen Balmer lines (Lintott et al. 2009; Keel et al. 2012). The [OIII] recombination time would be ~70 times shorter (Binette and Robinson 1987). The X-ray emission would then persist after the central continuum and [OIII] have faded. Using a more representative HI column density, and so a lower density, would lengthen the recombination time. This suggests that we might expect soft X-ray emission all the way back to the nucleus, and that "X-ray HVs" should be more common than [OIII] HVs.

In the thermal case, using the detected X-ray luminosity and a volume of 20 kpc length and 2kpc radius, we obtain gas densities of ~1 x $10^{-2}$ cm$^{-3}$ and a longer cooling time, ~$10^9$ yr. In this case, the X-ray emission would persist far longer than the optical emission. The $N_H$ estimate for the cold HI gas (see above) gives a density comparable (within a factor of ~3) with that of the X-ray gas, hence a much lower pressure. So, if the X-rays are thermal, the X-ray gas cannot be thermally confined by the HI clouds. However, the thermal expansion timescale for a ~kpc typical size D is $t_{th}$ ~ D / $c_s$ ~ 1kpc / 100 km s$^{-1}$ ~ $10^7$ yr. A $10^5$ yr old event would not then have dissipated yet.

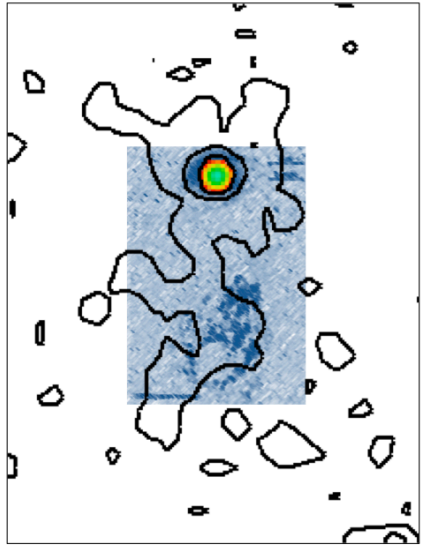

Figure 2.- Contours of the *Chandra* soft X-ray emission (from Figure 1) superimposed on the [OIII] *Hubble* image of Hanny's Voorwerp from Sartori et al. (2016), which also includes the peak position of the X-ray emission of IC 2497 (in colors).

3.2 Comparison with Photoionized Extended components of AGNs

The *Chandra* spectra of extended X-ray regions in nearby AGNs show both thermal and photoionization components, with photoionization dominating (e.g., Wang et al. 2011b; Paggi et al 2012; Fabbiano et al 2018a). Comparison of the [OIII] and soft (0.5 – 2 keV) X-ray luminosity of extended features in nearby AGNs shows a small range of ratios (Bianchi et al. 2006). Bianchi et al. (2006) demonstrated that higher [OIII]/X-ray ratios correspond to lower values of the ionization parameter U, defined as the ratio of the incident ionizing flux over the cloud density. With *Chandra*, cloud by cloud comparisons in NGC 4151 from HRC (Wang et al. 2009) and ACIS observations (Wang et al. 2011b) suggest consistent ratios, with values ~10 to 15, over a large range of radii (~20pc -~2000pc; see Figure 3).

The constant ratio of photoionized clouds in NGC 4151 is suggestive of emission from an expanding wind (Wang et al. 2009, 2011b; Figure 3). Although this is not necessarily a general feature. In NGC4151, the exception are clouds where thermal emission by radio jet driven shocks may occur, increasing the X-ray emission, and two inner clouds, where the ratios are significantly larger, implying relatively lower X-ray fluxes (Wang et al 2009; see left panel of Figure 3). Wang et al. (2009) suggest that these lower fluxes may result from the effect of intervening absorbers, covering the nuclear source. Alternatively, the [OIII] / soft X-ray ratio may be larger because of a higher density in these clouds as they are swept up by the outflow. Different localized regions in ESO 428-G014 also appear to be consistent with different ionization parameters (Fabbiano et al 2018b).

Using the integrated [OIII] flux for HV from Lintott et al. (2009) ~$3.2 \times 10^{-13}$ erg cm$^{-2}$ s$^{-1}$, and the (0.5-2 keV) luminosity of ~$1\times10^{40}$ erg s$^{-1}$ (for the parameters of Section 2), we estimate the [OIII]/soft X-ray luminosity ratio of the extended feature to be ~200.

Figure 3. compares this value with the clouds of NGC 4151 (Wang et al. 2009, 2011) and the average ratios of other nearby Seyferts (see the left panel of Fig. 3, from Wang et al. 2009). The HV ratio is higher (lower ionization parameter), than the average ratios shown in this figure, but perhaps comparable (given the uncertainties) with the highest values obtained for some of the NGC 4151 clouds.

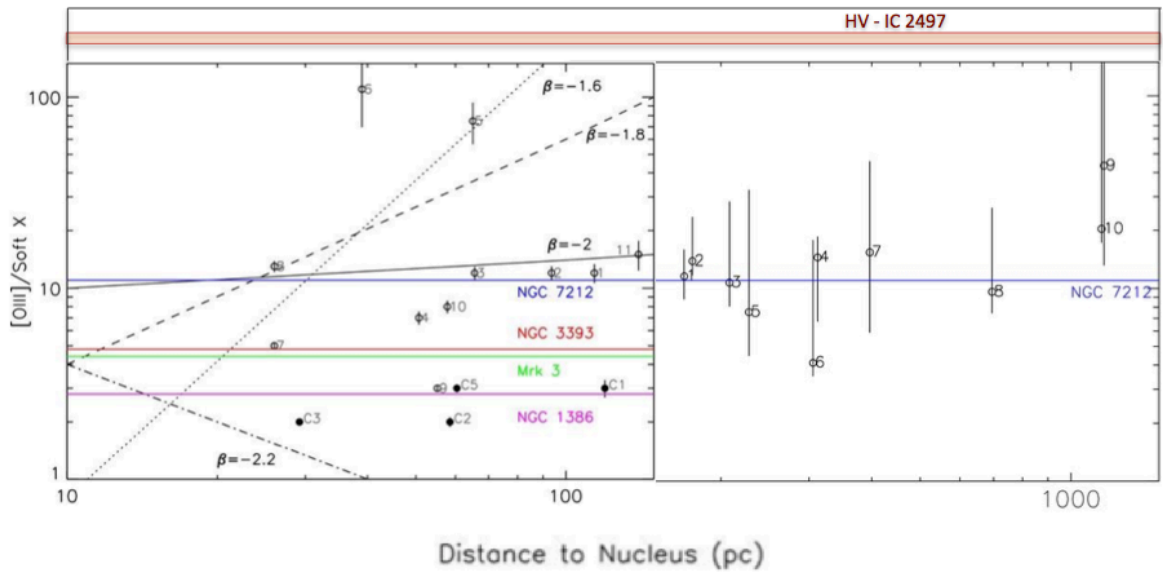

Figure 3. – This figure is an adaptation of Figure 6 of Wang et al. (2009; left panel) and Figure 15 of Wang et al. (2011b; right panel). We have added the estimated [OIII] to soft X-ray flux ratio for the HV feature (pink rectangle). While quite uncertain, this ratio is consistent only with the highest values measured from individual clouds in NGC 4151 (the circles with error bars). The filled circles identify the clouds in NGC 4151, where the X-ray emission is likely to contain an additional thermal component from the interaction with the radio jets (Wang et al 2009; Wang et al 2011a). As explained in Wang et al. (2009), the blue, red, green, and magenta lines indicate the [O III]/X-ray ratios for NGC 7212, NGC 3393, Mrk 3, and NGC 1386 (Bianchi et al. 2006), respectively. The dotted, dashed, solid, and dot-dashed lines are the CLOUDY model predicted values from Bianchi et al. (2006) for different radial density profiles: $n_e \propto r^\beta$ where $\beta = -1.6, -1.8, -2$, and $-2.2$, respectively.

Since the recombination time estimated for [OIII] is shorter than that estimated for the X-rays (Section 3.1), these high ratios cannot be due to the different time scales of 'fading' of a past quasar outburst. They may reflect differences in the incident ionizing flux and / or local cloud density relative to those observed in the majority of cases displayed in Figure 3. In contrast, the NGC 5252 relic has a low (~0.5-2) [OIII] / soft X-ray ratio (Dadina et al. 2010), comparable to the radio jet termination (thermal) regions in NGC 4151. Interestingly, Lintott et al. (2009) also find a low ionization parameter log U = -2.2 in the HV, from the [OII] 3727 to the [OIII] 5007 line ratio, although they also report high ionization lines (HeII 4616 and [NeV] 3346, 3426,) suggesting a complex emitting medium.

4. Conclusions

A new search of the *Chandra* images of IC 2497 for low surface brightness emission has led to the discovery of an elongated soft X-ray feature connecting the AGN with the emission line feature known as Hanny's Voorwerp. The X-ray feature is roughly spatially consistent with the radio continuum emission at 1.4 GHz (Jozsa et al. 2009). However, both the limited statistics of the X-ray data, and the relatively large radio beam of the Jozsa et al. data, impede a detailed comparison of these features. The present data are overall consistent with the spatial correspondence between X-ray, optical emission line, and radio features detected in nearby AGN (see e.g., ESO 428-G014, Fabbiano et al 2018b).

The average (0.5 – 2. keV) X-ray luminosity ~1 x $10^{40}$ erg $s^{-1}$, compared to the inferred [OIII] luminosity L[OIII]~4-14 x $10^{41}$ erg $s^{-1}$, is consistent with the highest values measured in some of the clouds of NGC 4151 (Wang et al. 2009; 2011b), although higher than average values for other nearby Seyferts. Overall, given the uncertainties, Hanny's Voorwerp could be a feature consistent with the ionization cone emission of nearby AGNs. If both [OIII] and soft X-ray emission are due to photoionization, the X-ray recombination time is ~2 x $10^7$ yr, longer than the [OIII] recombination time (a few 1000 years or less). This suggest that extended soft X-ray components may be a better diagnostic of overall activity, while detection of an [OIII] HV would point to a time-limited activity burst.

Future deeper high-resolution X-ray and radio observations will be needed to explore this extended emission at higher resolution and statistical significance. Unreasonably long exposure times would be required to pursue this investigation with *Chandra*, because of the deterioration of the ACIS response at the lower energies. With *Chandra* we could achieve comparable data quality on HV as the *Einstein Observatory* could yield for the extended emission of NGC 4151 (Elvis et al. 1983). To reach the level of detail *Chandra* gives for NGC 4151 (e.g., Wang et al 2011a, b) for HV will require *Lynx*. The proposed *Lynx* mission would give 1000 counts in ~75 ksec, sufficient for detailed mapping and spectra.

We thank Junfeng Wang for useful discussions and Peter Maksym for comments on the manuscript. We retrieved data from the *Chandra* Data Archive and the NASA-IPAC Extragalactic Database (NED). For the data analysis we used the CIAO toolbox, DS9, and on-line software tools developed by the *Chandra* X-ray Center (CXC). This work was partially supported by NASA contract NAS8- 03060 (CXC). This work was performed in part at the Aspen Center for Physics, which is supported by National Science Foundation grant PHY-1607611.